\documentclass[
 reprint,
superscriptaddress,
amsmath,
amssymb,
aps,
pra,
floatfix
]{revtex4-2}

\usepackage{graphicx}
\usepackage{titlesec}
\titlespacing{\section}{0pt}{4ex}{2ex}
\titleformat{\section}{\bfseries\raggedright}{\thesection.}{0.5em}{}
\usepackage[skip=0pt, indent=20pt]{parskip}
\usepackage{float}
\usepackage{dcolumn}
\usepackage{hyperref}
\usepackage{amssymb}
\usepackage{chemformula}
\usepackage[utf8]{inputenc}
\usepackage[T1]{fontenc}
\usepackage[mathlines]{lineno}
\usepackage{stackengine}[2013-10-15]

\begin{document} 


\title{Epitaxially-stabilized growth of wüstite FeO on 4H-SiC}

\author{Faisal Kimbugwe}
\author{Marzieh Baan} 
\author{Alexandra Fonseca Montenegro}
\affiliation{Department of Materials Science and Engineering} 
\author{Roberto C. Myers} 
\affiliation{Department of Materials Science and Engineering} 
\affiliation{Department of Electrical and Computer Engineering} 
\affiliation{Department of Physics} 
\author{Tyler J. Grassman}
\email{grassman.5@osu.edu} 
\affiliation{Department of Materials Science and Engineering} 
\affiliation{Department of Electrical and Computer Engineering} 
\affiliation{Center for Electron Microscopy and Analysis \\
The Ohio State University, Columbus OH 43210, USA}

\begin{abstract}
\vspace{0.3cm}
Iron(II) monoxide (FeO) is thermodynamically stable in the halite (wüstite) structure only at elevated temperatures in a typically non-stoichiometric, Fe-deficient, Fe$_{1-z}$O form that tends to phase separate and/or transform into metallic $\alpha$-Fe and magnetite Fe$_3$O$_4$ at ambient conditions. Here we report on the successful growth of up to 180 nm thick $(111)$-oriented FeO heteroepitaxial films on slightly lattice-matched 4H-SiC$(0001)$ using molecular beam epitaxy (MBE). The films have flat, terraced surfaces with tall multi-layer steps. X-ray diffraction (XRD), high-resolution scanning transmission electron microscopy (S/TEM), energy-dispersive X-ray spectroscopy (EDS), and core-level electron energy loss spectroscopy (EELS) collectively confirm the epilayer as phase-pure wüstite FeO, with atomically sharp FeO/SiC interfaces. The films are found to exhibit a slight misfit strain-induced rhombohedral distortion that does not appear to vary over the range of thicknesses examined. These results demonstrate the power of epitaxial stabilization for integrating a thermodynamically unstable, yet functionally interesting material with a commercially available and technologically important semiconductor platform.
\end{abstract}

\maketitle


\section{INTRODUCTION}

Of the various phases of iron oxide, FeO (wüstite) is unique in that it exhibits the same simple halite structure (i.e. rocksalt, or NaCl) as many transition metal monoxides (e.g., MgO, NiO, MnO, CoO, CrO). However, unlike these other compounds, FeO is thermodynamically unstable under ambient conditions, with a preference toward decomposing into the stable mixed Fe$^{2+/3+}$ inverse spinel Fe$_3$O$_4$ (magnetite) and metallic Fe (bcc) phases. To circumvent decomposition, metastable bulk iron(II) monoxide specimens are typically obtained by fast quenching to ambient, but even at elevated temperatures the pure stoichiometric form is still unstable, resulting in generally non-stoichiometric compositions, Fe$_{1-z}$O, with Fe(II) deficiencies spanning $0 < \textit{z} \leq 0.16$ \cite{gronvold_heat_1993, non_stoic_Zeng, Berthon_wustite}. Standard charge compensation models dictate that the Fe$^{2+}$ vacancies must be accommodated through the introduction of Fe$^{3+}$ "impurities," which can reside at either the halite octahedral sites or the tetrahedral interstitials, yielding a continuous pathway toward the formation of the energetically preferred magnetite phase. 

Nevertheless, non-equilibrium conditions can often be employed to stabilize metastable structures, such as in the formation of an amorphous (noncrystalline) configuration via physical vapor deposition onto a cold substrate. Alternatively, epitaxial synthesis, which typically occurs far from equilibrium, can provide a kinetic pathway to stabilization by establishing a bonding template for the deposited adatoms toward a particular structure. A demonstrative and relevant example is the relatively new monoclinic form of iron oxide, $\mu$-Fe$_2$O$_3$, which can be stabilized by the monoclinic structure of $\beta$-Ga$_2$O$_3$, enabling facile thin-film growth of the otherwise unstable material via molecular beam epitaxy (MBE) \cite{jamison_ferromagnetic_2019, hettiaratchy_interface-induced_2020}. A similar process is proposed here for FeO, whereby epitaxial stabilization by lattice near-matching between the atomic planes of $(111)$-oriented FeO and hexagonal 4H-SiC$(0001)$, in conjunction with the ability to precisely control the oxidation environment, may provide an adequate kinetic pressure to stabilize the phase under ambient conditions.

\begin{figure*}[t!]
  \includegraphics[width=1\linewidth]{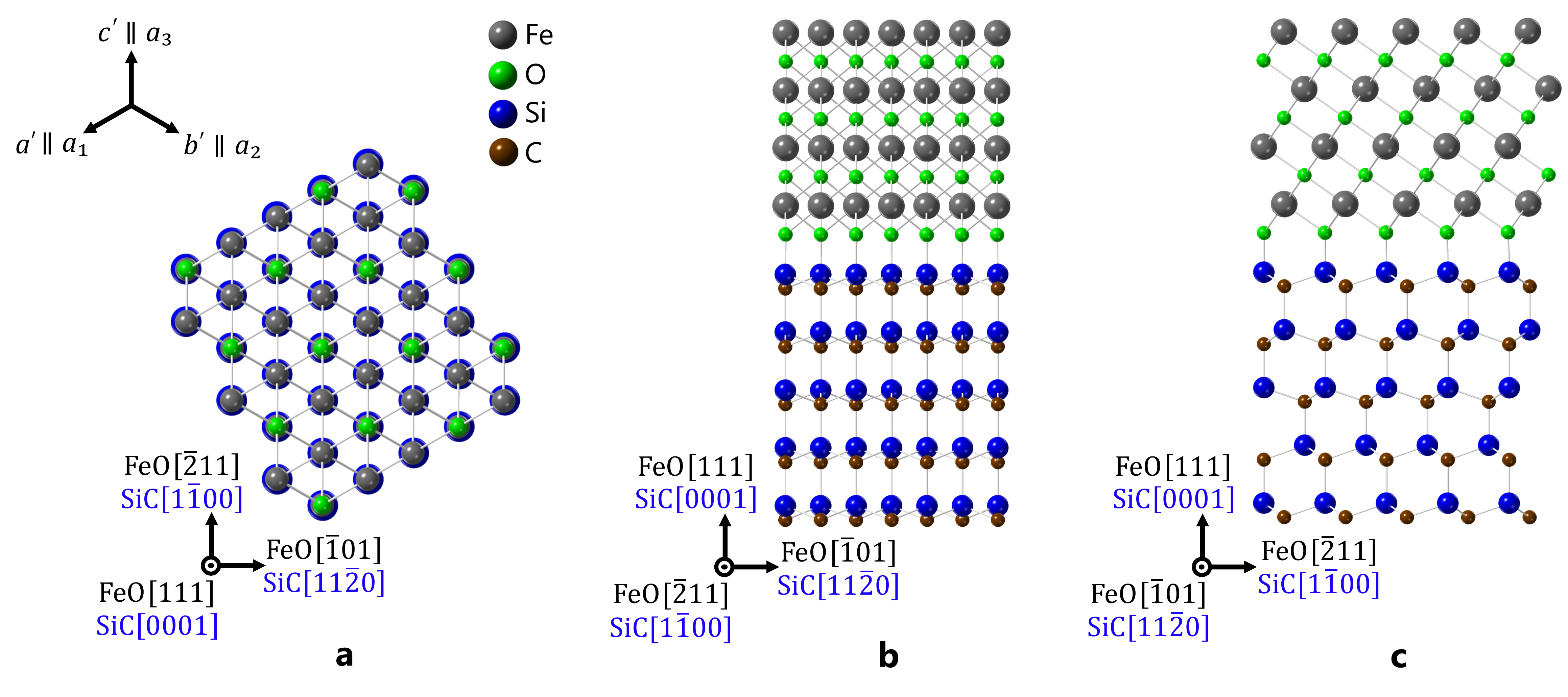} 
 \caption{Schematic of the proposed FeO/SiC heteroepitaxial system as viewed in the (a) FeO$[111]$ / SiC$[0001]$, (b) $[\bar{2}11]$ / $[1\bar{1}00]$, and (c) $[\bar{1}01]$ / $[11\bar{2}0]$ directions. The interfacial coincidence lattice matching is also indicated in (a).} 
  \label{fig1} 
\end{figure*}

Oxide thin films are typically grown on oxide substrates, simplifying the interface by providing a unified isovalent anionic (oxygen) sublattice. Additionally, the oxidative environment required for film growth can potentially lead to undesired reactions with a non-oxide substrate surface, introducing interface phases that disrupt the epitaxy. Indeed, the most successful efforts toward FeO epitaxy to date have been based on oxide substrates. Rubio-Zuazo \textit{et al.} reported nearly 8 nm thick, single-phase, epitaxial $(001)$-oriented FeO on highly lattice-mismatched ($f\approx 10$\%) SrTiO$_3$ via pulsed laser deposition (PLD).\cite{rubio-zuazo_incommensurate_2015, rubio-zuazo_electronic_2018}. Similarly, Ruby \textit{et al.} and Kozioł-Rachwał \textit{et al.} both report epitaxial wustite-phase FeO$(001)$ thin films via growth on MgO \cite{Ruby_MGO} \cite{Ultra_thin_feO_MGO, Ultra_thin_feO_MGO_part1}. However, despite the much smaller (but still relatively large) mismatch of only $f = 2.8\%$, the reported film thicknesses were still $\leq 6$ nm.

Ultrathin FeO films have also been reported for growth on relatively oxidation-resistant noble metal substrates. For example, Pt$(111)$ was used to study the initial growth and phase transformations of iron oxides \cite{weiss_surface_2002}, wherein pure FeO was stable only up to 2.5 monolayers, after which Fe$_3$O$_4$ islands began to form. Similar FeO phase instability was observed for MBE-based growth on Ag$(100)$ \cite{Ag_Abreu}, where post-deposition vacuum annealing was used to convert the Fe$_3$O$_4$ to FeO. However, thicker epilayers inevitably yielded increasing Fe$_3$O$_4$ content, regardless of post-deposition processing.  

Given the large lattice and/or chemical mismatches in the aforementioned epitaxial heterostructures, one would expect the resulting misfit strains and interfacial energies to have a substantial impact on the growth morphology (e.g., Stranski-Krastanov) and/or extended defect populations very early within the growth, precluding any potential epitaxial stabilization effects. Any phase stabilization via an "epitaxial kinetic pressure" will likely require a more closely lattice-matched substrate. To this end, we identify $(0001)$-oriented 4H-SiC as a near-matched, commercially mature, and readily available high-quality substrate for this purpose. Here, the most favorable coincident lattice alignment is FeO$(111) || $ SiC$(0001)$, where the orthogonal projections of the FeO lattice constants $(a', b', c')$, are parallel to the SiC basal constants, $(a_1,a_2,a_3)$, as depicted in Figure 1(a). The misfit strain for the as-described (111)-oriented cubic film, with lattice constant $a_f=a_{FeO}$, grown on a (0001)-oriented hexagonal, with lattice constant $a_s=a_{SiC}$, is given by

\begin{equation}
\label{eq:misfit_strain} f = \frac{a_{s} - a_{f}/\sqrt{2}}{a_{f}/\sqrt{2}}= \frac{a_{SiC} - a_{FeO}/\sqrt{2}}{a_{FeO}/\sqrt{2}}.
\end{equation}

Studies on quenched bulk wüstite specimens \cite{gronvold_heat_1993, Berthon_wustite} have established a linear relationship for the equilibrium cubic Fe$_{1-z}$O lattice constant, $a_{FeO}$, with respect to the Fe vacancy fraction, $z$ (at least up to about $z=0.12$), such that

\begin{equation}
\label{eq:lattice_constant} a_{FeO} \approx 4.332\text{\AA} - 0.43z.
\end{equation}

The corresponding variation in $f$ for FeO$(111)$ / SiC$(0001)$ epitaxy, with $a_{SiC}=3.073$ \AA, is then expected to range from a relatively small mismatch of only $\sim 0.31\%$ at fully stoichiometric FeO ($z=0$) to $1.5\%$ at $z=0.12$. Assuming adequate control over the interfacial chemistry and Fe$_{1-z}$O stoichiometry, the relatively small lattice mismatch within this system is expected to support the stabilized growth of pure wüstite thin films.

To this end, we report here on the successful growth of up to 180 nm thick wüstite phase FeO$(111)$ films on 4H-SiC$(0001)$ using molecular beam epitaxy (MBE). This was achieved through a combination of pre-growth substrate preparation to establish the desired heterovalent interfacial chemistry, growth conditions optimized to lie within the relatively narrow window for FeO stability, and rapid quenching to room temperature. Characterization of the resultant epilayers via a range of structural and chemical / spectroscopic modalities confirms the crystal phase as pure wüstite FeO, over an order of magnitude thicker than any previously reported.

\section{EPITAXIAL METHODS}

All growths presented herein were performed within an Applied EPI Gen II MBE reactor customized for oxide thin-film deposition. The chamber maintains a typical pre-growth background pressure of about $5\times10^{-10}$ torr. A high-temperature effusion cell with a BeO crucible was used for the Fe source ($99.98\%$ purity). An Fe beam equivalent pressure (BEP) of $6.4\times10^{-8}$ Torr was used for all growths, yielding an approximate FeO growth rate (based on post-growth glancing X-ray reflectance thickness measurements) of $\sim 5.1$ nm/min. Ultra-high purity (5N) O$_2$ was supplied through an unlit plasma source and controlled using a high-precision ultra-high vacuum leak valve. To avoid ion gauge filament damage, the molecular O$_2$ flux was not directly measured via BEP, but rather with the chamber gauge outside the internal cryo shroud; the actual effective pressure at the sample surface during growth is estimated to be approximately one order of magnitude higher. Typical O$_2$ chamber pressure ($p_{O_2}$) for these growths was $2.2\times10^{-5}$ torr.

Substrates employed for this work were 150 mm diameter, $(0001)$-oriented, Si-face 4H-SiC ("production grade" from Wolfspeed). The base substrates are highly-conductive \textit{n}-type (N-doped, 0.015 - 0.028 $\Omega$-cm), with 5 $\mu$m of $N_D=1\times10^{16} cm^{-3}$ and 0.5 $\mu$m $N_D=1\times10^{18} cm^{-3}$ \textit{n}-type homoepitaxy. The wafers were intentionally misoriented with a 4.0$^\circ$ offcut toward $[11\bar{2}0]$. Before any growth or pre-growth preparation, the SiC wafers were diced into $1 \times 1$ cm squares, using photoresist and dicing tape to protect the surfaces. As previously noted, the use of non-oxide substrates can provide a challenge for oxide epitaxy. To prepare the substrates for growth, the amorphous native oxide was removed using a 10-minute etch in 10$\%$ HF solution (5mL HF: 45mL DI H$_2$O) followed by a 1-minute DI rinse. Unlike Si, where HF-based deoxidation yields a hydrophobic H-terminated surface, Si-face SiC results in an air-stable hydroxyl (OH) surface termination \cite{OH_termination}, which serves as a nominally ideal starting surface for subsequent oxide growth.

After HF etching, the substrates were loaded into the sample holder using a Si backing wafer to improve absorption of the heater's radiation. The sample block was loaded into the MBE chamber through a load-lock assembly and thermally outgassed at 400$^\circ$C to remove any residual moisture or atmospheric adsorbates from the surface. The substrate was heated to the target growth temperature ($T_{sub}$) of 750$^\circ$C, based on the heater thermocouple reading, while facing away from the effusion source flanges to avoid excessive exposure and potential oxidation of the pristine starting surface. The O$_2$ leak valve was then opened and adjusted to stabilize the background growth chamber pressure ($p_{O_2}$). Upon stabilization of the target growth conditions, the substrate was rotated to the growth position, and the Fe effusion cell shutter was opened to commence deposition. For this combination of Fe flux, $p_{O_2}$, and $T_{sub}$, all samples grown were found to yield phase pure wüstite FeO epilayers; data is provided herein on a subset of samples with nominal epilayer thicknesses of 26 nm (Sample A), 38 nm (B), 80 nm (C), and 180 nm (D). Following completion of growth, the Fe and O$_2$ shutters were closed and the sample was allowed to free-fall to room temperature as an approximation to fast quenching.

\section{RESULTS AND DISCUSSION}

\begin{figure}[h!]
  \includegraphics[width=\columnwidth]{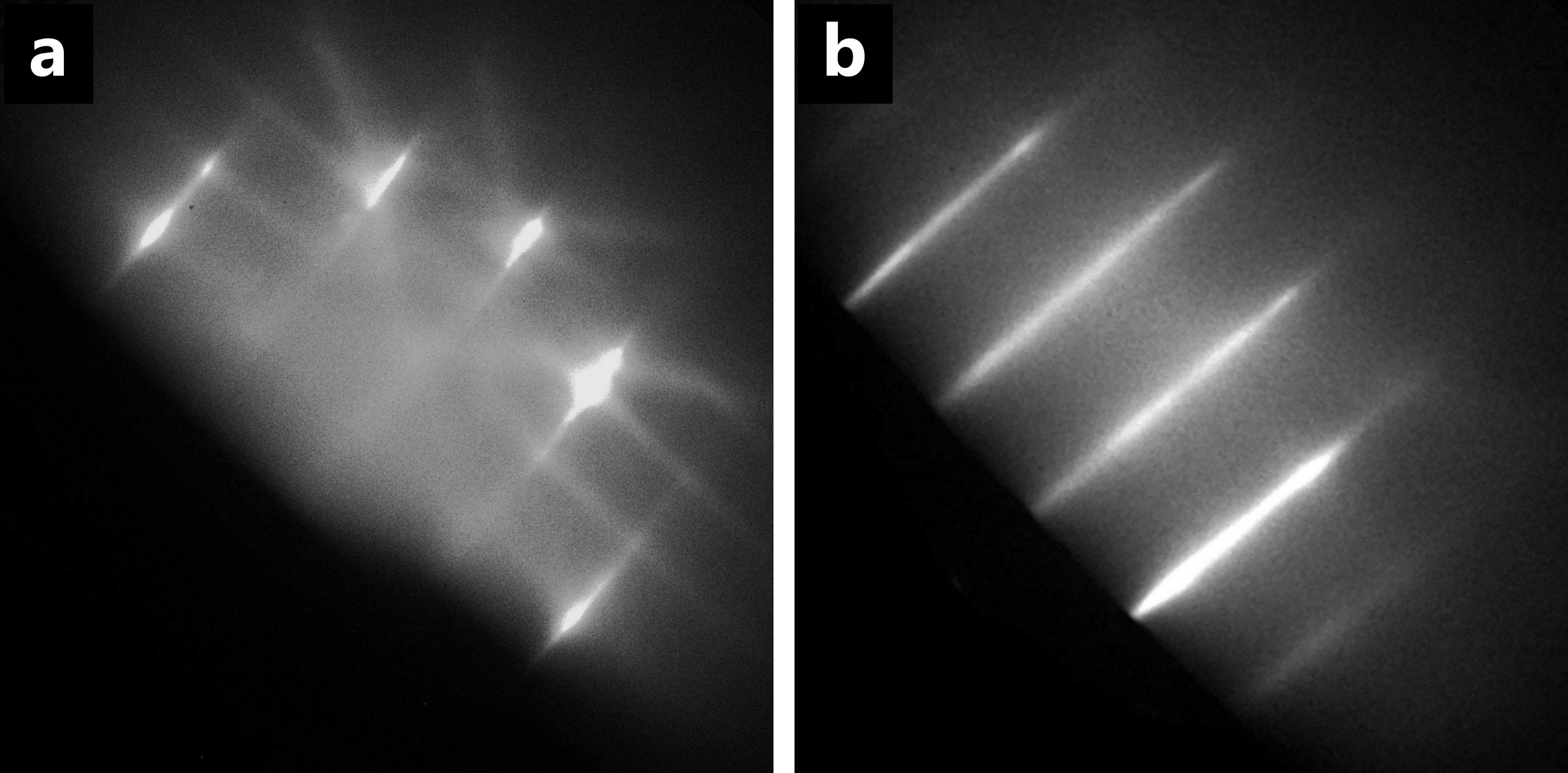} 
\caption{RHEED images captured of the (a) bare SiC$(0001)$ surface immediately prior to growth initiation and (b) after 1 minute of FeO growth. Both images were taken along the SiC$[11\bar{2}0]$ direction at 750$^\circ$C.} 
  \label{fig2} 
\end{figure}

In situ growth monitoring was performed using reflective high-energy electron diffraction (RHEED). Fig. \ref{fig2}(a) shows a sharp, streaky 1$\times$1 RHEED pattern, taken along the $[11\bar{2}0]$ direction, from the OH-terminated 4H-SiC(0001) substrate immediately prior to FeO growth initiation \cite{RHEED_SiC_Kerrigan}, confirming both removal of the native SiO$_2$ and the existence of a reasonably smooth starting surface. The orthogonal SiC Kikuchi lines quickly subside upon the start of FeO deposition, and a strong, streaky 1$\times$1 pattern emerges and is retained throughout the growth. Fig. \ref{fig2}(b) presents an example RHEED pattern image taken after about 1 minute of FeO growth. The line spacing is similar to that of the SiC, consistent with that expected from an unreconstructed wüstite FeO(111). 

\begin{figure}[t!]
  \includegraphics[width=\columnwidth]{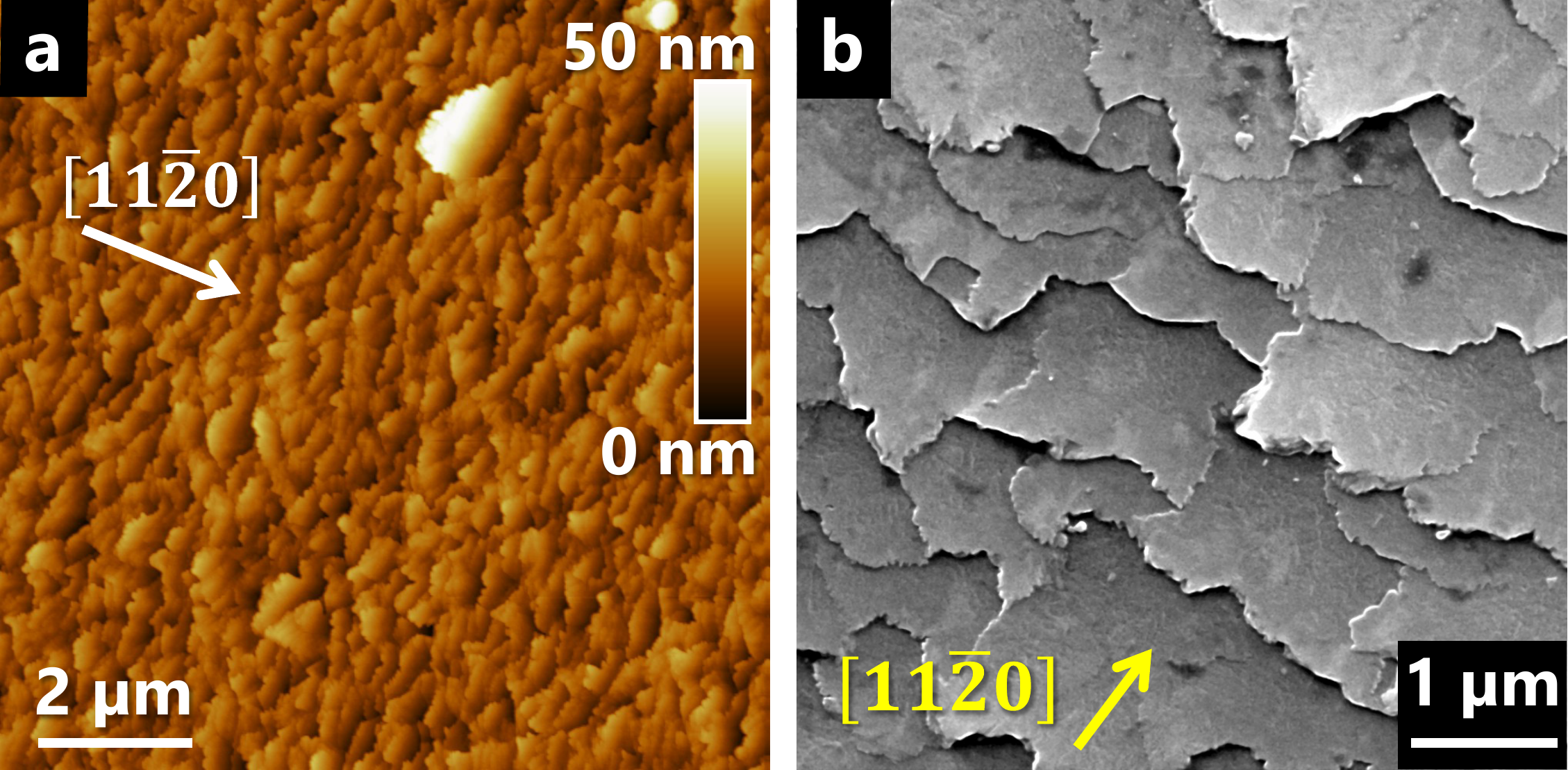} 
\caption{(a) AFM image of an 80 nm FeO thin film (Sample C) grown on SiC$(0001)$. (b) SEM image of a 180 nm thick FeO film (Sample D) grown on SiC$(0001)$.} 
  \label{fig3} 
\end{figure}

Characterization of the epilayer surface morphology was performed via atomic force microscopy (AFM), using a Bruker Icon AFM in tapping mode, and scanning electron microscopy (SEM), using a Thermo Scientific Quattro SEM in standard secondary electron mode. Figure \ref{fig3}(a) shows a representative AFM image of the FeO epilayer surface, taken from Sample C. The surface was found to possess a relatively high density of steps along the offcut direction, SiC$[11\bar{2}0]$ / FeO$[\bar{1}01]$, with otherwise flat and smooth terraces. The steps have an average height of 18.3 $\pm$ 2.1 nm, with an average terrace width of 300 $\pm$ 60 nm; there is a relatively large variance in both step height and width, making precise quantification somewhat challenging. These values correspond to a geometric misorientation of $\sim 3.5^\circ$, reasonably consistent with the $4.0^\circ$ specified wafer offcut. The tall steps, equivalent to an average of $\sim 42$ unit cells, with sharp vertical edges, indicate substantial step-bunching and suggest a high degree of stability of the surface-orthogonal FeO$\{1\bar{1}0\}$ crystal planes. Higher magnification SEM imaging, collected on Sample D and shown in Fig. \ref{fig3}(b), confirms the stepped nature of the surface and flatness of the terraces.

\begin{figure}[t!]
  \includegraphics[width=0.9\columnwidth]{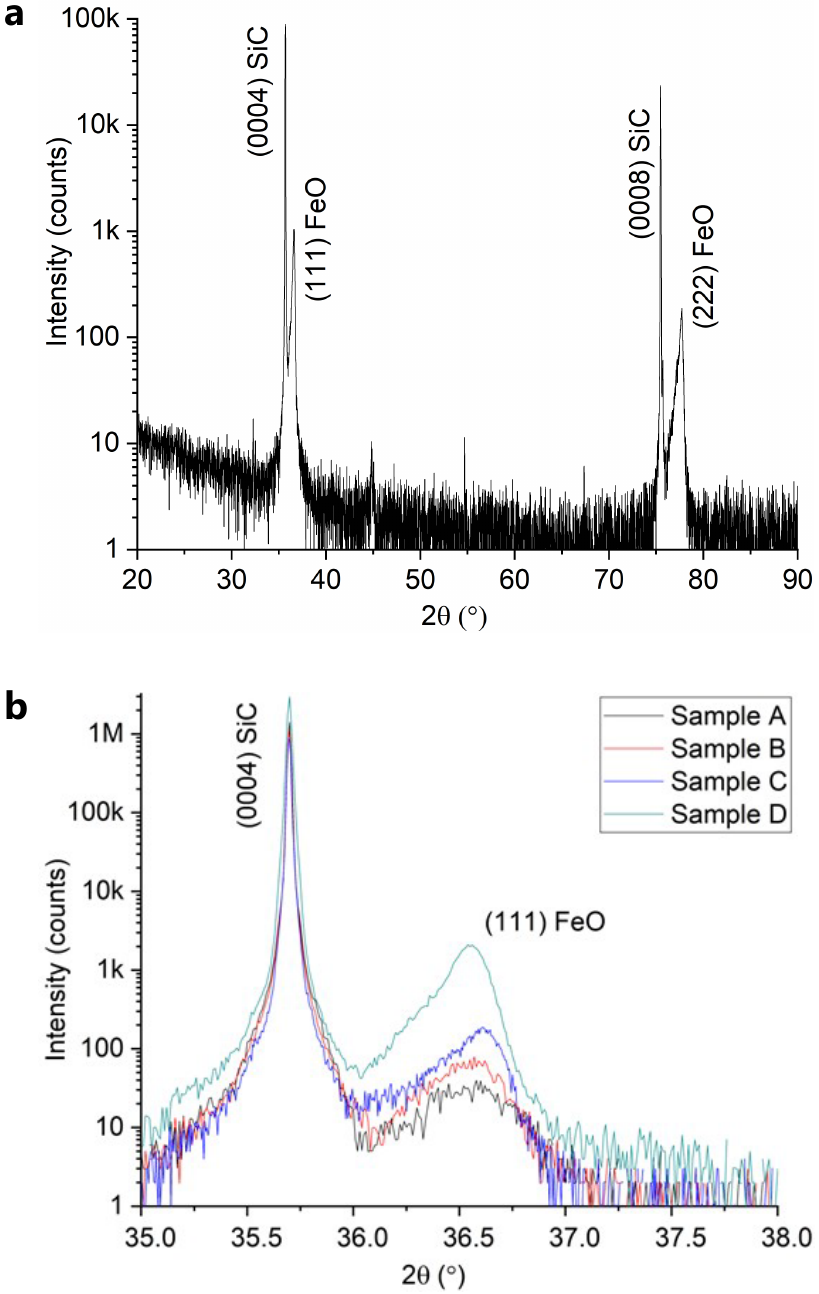} 
\caption{(a) A representative long-range $\theta-2\theta$ XRD scan of Sample C, taken along the SiC$[11\bar{2}0]$ zone, indicating wüstite FeO as the only non-substrate phase present. (b) HRXRD scans collected from Samples A, B, C, and D around the $(0004)$ SiC peak. Of any potential iron oxide phases, only the near-matched $(111)$ FeO is observed in all samples.} 
  \label{fig4}
\end{figure}

To confirm the crystal structure of the epilayer, high-resolution X-ray diffraction (XRD) $\theta-2\theta$ scans were collected using a Bruker D8 diffractometer. Figure \ref{fig4}(a) presents an example long-range scan taken from Sample C along the SiC$[11\bar{2}0]$ zone, where no appreciable peaks other than those expected for the epitaxial FeO$(111)$ / 4H-SiC$(0001)$ system are visible; indeed, this is the case for all samples produced using the noted growth conditions. Figure \ref{fig4}(b) presents higher resolution scans from around the $(0004)$SiC region comparing Samples A \textendash{} D; note that the $(0008)$SiC oriented scans are effectively identical and thus not shown here. As before, only a single peak is observed near the substrate, located about where $(111)$FeO is expected; magnetite (Fe$_3$O$_4$), if present, would be expected at a $2\theta$ angle of $\sim 37.5^\circ$. The nominal $(111)$FeO peaks are relatively broad, suggesting some degree of nonuniformity, likely related to strain and/or the presence of significant Fe vacancy populations, in addition to the expected thin-film broadening. Analysis of the peak positions based on a cubic unit cell suggest lattice constants ranging around 4.25 \textendash{} 4.27 \AA{}, substantially smaller than the vacancy-free lattice constant of 4.332 \AA{}; such values would suggest an unrealistic vacancy concentration on the order of $\sim20\%$ (from Eq. \ref{eq:lattice_constant}), notably higher than anything previously reported. However, because this is a lattice-mismatched heteroepitaxial system, it is not possible to isolate the impacts of both misfit strain and composition (i.e., vacancy fraction) using only simple $\theta-2\theta$ HRXRD data. 

\begin{figure}[t!]
  \includegraphics[width=\columnwidth]{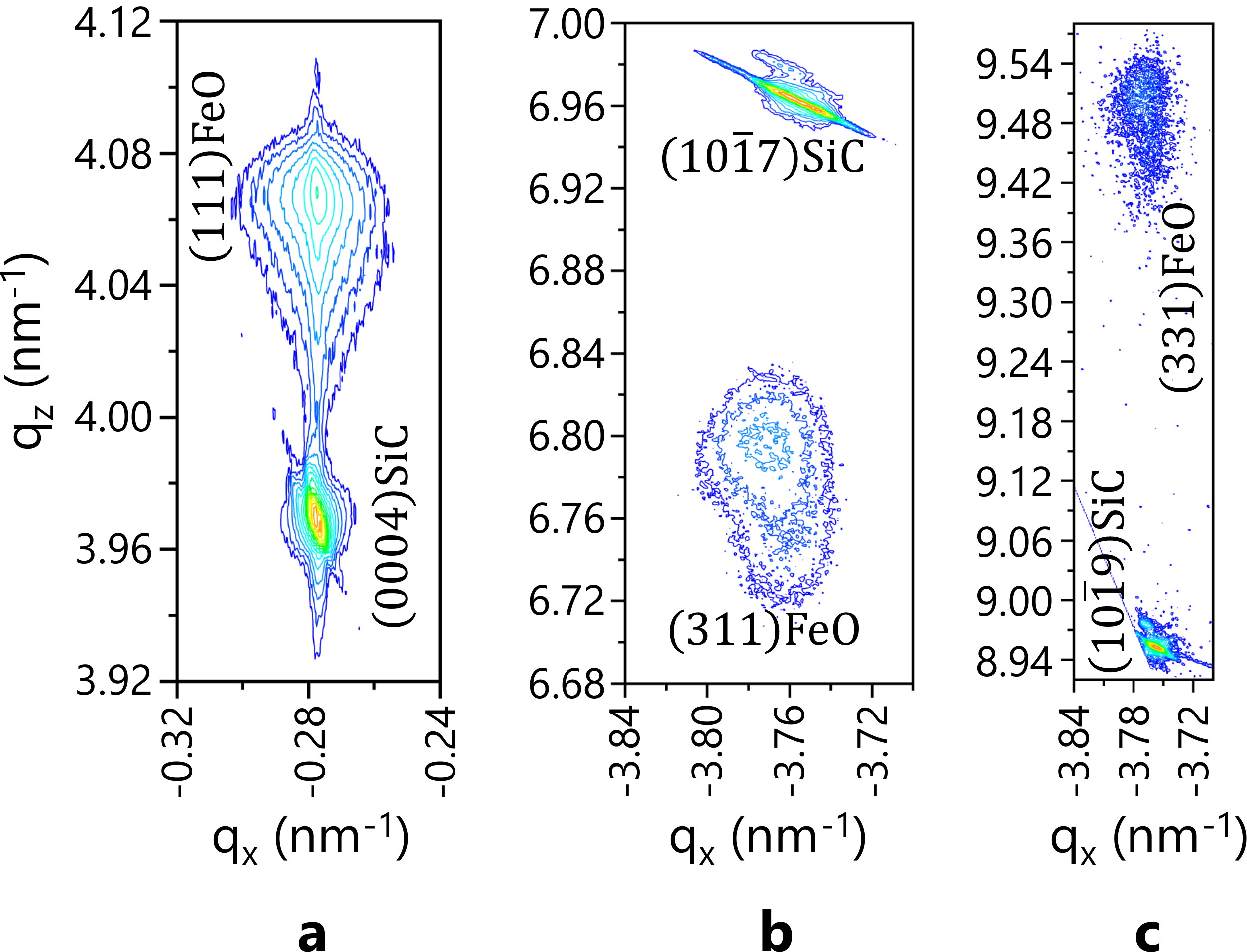} 
\caption{Representative XRD reciprocal space maps, collected from Sample C, taken around the (a) symmetric $(0004)$SiC / $(111)$FeO, (b) asymmetric $(10\bar17)$SiC / $(311)$FeO, and (c) asymmetric $(10\bar19)$SiC / $(331)$FeO diffraction peaks.} 
  \label{fig5} 
\end{figure}

To further investigate the epitaxial relationship, crystalline quality, and strain state of the FeO films, XRD reciprocal space maps (RSMs) were collected around the symmetric $(0004)$SiC and glancing incidence asymmetric $(10\bar{1}7)$SiC and $(10\bar{1}9)$SiC conditions, corresponding to the nominal $(111)$FeO, $(311)$FeO, and $(331)$FeO diffraction conditions, respectively. Representative RSMs of each, taken from Sample C, are presented in Figure \ref{fig5}. As seen in Fig. \ref{fig5}(a), the $(111)$FeO and $(0004)$SiC peaks are precisely aligned on the q$_x$ axis, confirming the epitaxial orientation. Moreover, consistent with the HRXRD scans, the $(111)$FeO peak is elongated in the q$_z$ axis, likely due to the aforementioned strain and/or compositional non-uniformity. Further, the peak is relatively broad in the q$_x$ (4$\times$ larger than the substrate peak), suggesting a non-negligible degree of mosaicity or atomic-scale disorder. The latter observation is relatively unsurprising for a material that is expected to possess a significant concentration of point defects, especially given the propensity for clustering \cite{non_stoic_Zeng}, as well as some anticipated microstructural content related to the epitaxial lattice mismatch. The q$_x$ alignment between the substrate and epilayer peaks in the asymmetric RSMs provides additional confirmation of close in-plane matching between FeO and SiC. Consistent with the cubic unit cell analysis from the HRXRD $\theta-2\theta$ scans, an unrealistically small lattice constant for the FeO epilayer is also obtained here from the individual RSMs when using the same cubic assumption. However, the lattice constants extracted from the asymmetric $(311)$ and $(331)$ peaks, Figs. \ref{fig5}(b,c),  do yield $\sim 0.5\%$ larger lattice constants than determined from the $(111)$, indicating an in-plane versus out-of-plane structural distortion consistent with the tensile misfit at the FeO/SiC interface.  

More importantly, the combination of symmetric and asymmetric RSMs allows for the deconvolution of in-plane and out-of-plane lattice spacing, and thus a more accurate determination of the strain-induced distortions of the crystal structure. As indicated in Fig. \ref{fig1}(a), the three lattice vectors for the $(111)$-oriented halite unit cell of FeO have equal projections along the equivalent SiC lattice vectors, and are thus under nominally equal magnitudes of tensile strain. The 4$^\circ$ wafer offcut does create a slight geometric anisotropy, but it is small enough to be considered effectively negligible. The symmetric in-plane tensile strain on the $(111)$-oriented cubic unit cell would be expected to produce a commensurate compression along the $[111]$ direction, thereby producing an overall rhombohedral distortion. Employing all three RSMs collected from Sample C within a non-linear least squares fit to a rhombohedral unit cell model reveals a lattice constant of $a_{FeO}$ $= 4.306\pm 0.004$ \AA{} and lattice angle $\alpha = 90.78^\circ \pm 0.06^\circ$; the same procedure for Sample D yields identical values. Indeed, these parameters give an in-plane Fe-Fe (or O-O) distance of $3.066$ \AA{}, a nearly exact match to the expected in-plane Si-Si distance of the SiC at $3.073$ \AA{} (i.e., the $a_{4H-SiC}$ lattice constant). For a purely cubic cell, this FeO lattice constant provides an Fe-Fe / O-O distance of only $3.045$ \AA; that is, a lattice mismatch of $0.68\%$.

The excellent fit, with minimal errors, provides a high degree of confidence for the rhombohedral distortion model. While this is a relatively small distortion (a difference of only $\sim0.78^\circ$), it easily explains the discrepancies encountered when assuming a purely cubic cell and fully accounts for the in-plane matching indicated by the RSMs. The rhombohedral distortion also points to the presence of unrelaxed (elastic) misfit strain within the film. In fact, it is surprising to find identical lattice parameters in both samples C and D, which differ in thickness by over 2$\times$, suggesting that either the barriers to the generation of microstructure-based strain relaxation, like the formation of misfit dislocations, may be relatively high or that the rhombohedral distortion is particularly effective at accommodating these moderate strain levels. It is unknown whether the established trends with respect to vacancy concentration are still applicable in this strain-distorted case. However, if we assume that they are, then a relaxed equilibrium cubic lattice constant for FeO of ${a}_{FeO} = 4.306$ \AA{} would correspond to a vacancy fraction of $z = 5.9\% \pm 1.7\%$, which is well within the ranges reported for quenched bulk specimens. 

To directly examine the crystal structure of the FeO epilayer and FeO/SiC interface, scanning transmission electron microscopy (S/TEM) was employed. After depositing a thin Ir surface layer to protect the epilayer, a cross-sectional lamella, aligned to enable imaging on the $[1\bar{1}00]$SiC / $[\bar{2}11]$FeO zone axis, was extracted from Sample D using an FEI Helios 600 focused ion beam (FIB) system. The specimen was thinned down to $\sim 50  nm$, being careful to use sufficiently low ion beam energies and currents to minimize damage. The thinned specimen then received a final low-energy Ar$^+$ polish in a Fishione 1040 Nanomill to remove any remaining surface amorphization. Subsequent S/TEM imaging and spectroscopic analysis was performed in a Thermo Scientific Themis Z aberration-corrected microscope operating at an accelerating voltage of 200 kV. The microscope was equipped with a Super-X energy dispersive X-ray spectroscopy (EDS) system for high sensitivity elemental analysis and a Gatan Imaging Filter (GIF) electron energy loss spectroscopy (EELS) system.

\begin{figure*}[ht!]
  \centering
  \includegraphics[width=0.9\textwidth]{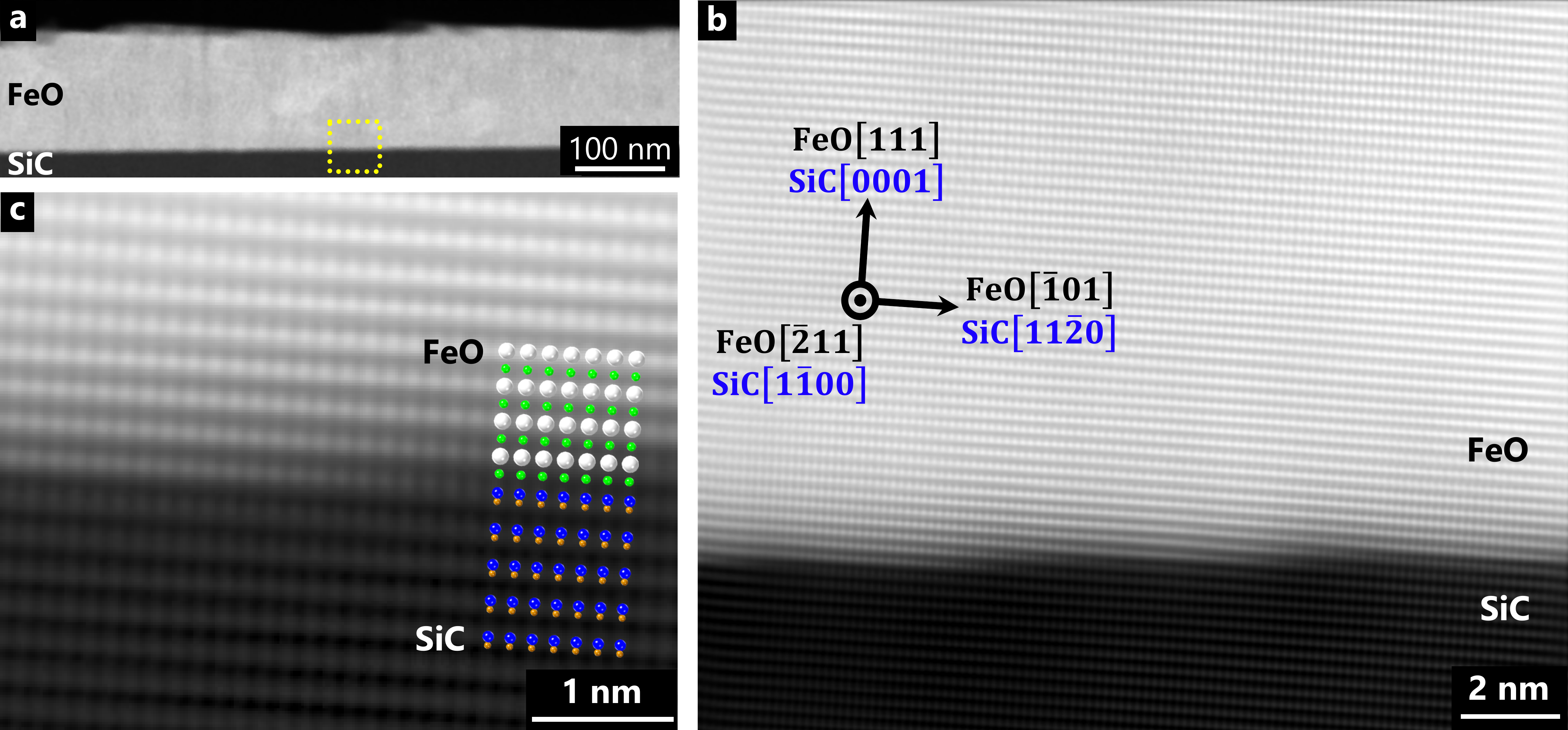} 
\caption{(a) Low-magnification HAADF S/TEM image of Sample D, taken at the $[1\bar{1}00]$ SiC zone axis. (b) Atomically-resolved image showing the sharp FeO/SiC interface and wüstite crystal structure. (c) High-magnification image of the FeO/SiC interface with crystal structure model overlay.} 
  \label{fig6} 
\end{figure*}

Figure \ref{fig6}(a) presents a low-magnification high-angle annular dark-field (HAADF) image of the FeO/SiC epitaxial heterostructure. The FeO and SiC are easily distinguished due to the large difference in atomic number (Z-contrast) for Fe versus Si. The tall surface steps observed in the AFM and SEM imaging are also visible at the top of the lamella. Fig. \ref{fig6}(b) and \ref{fig6}(c) provide representative high-magnification, atomically-resolved images of the FeO/SiC interface taken from the region indicated in \ref{fig6}(a) by the yellow box. Here we can see a crisp, atomically sharp interface with no apparent intermixing, consistent with the nominal Si-O-Fe interfacial bridging structure predicted in Fig. \ref{fig1}(b). Double Si-C bilayer surface steps resulting from the 4$^\circ$ wafer offcut are also visible; the slight blur at their edges is because they are not perfectly straight along the $[1\bar{1}00]$SiC viewing direction. The well-ordered structure of the epilayer is an exact match to that expected for the wüstite FeO structure, as depicted in Fig. \ref{fig6}(c); other iron oxide structures (e.g. magnetite, Fe$_3$O$_4$) would show drastically different patterns. As indicated by the asymmetric RSMs, the interface is fully coherent and atomically aligned.

\begin{figure}[h!]
  \includegraphics[width=\columnwidth]{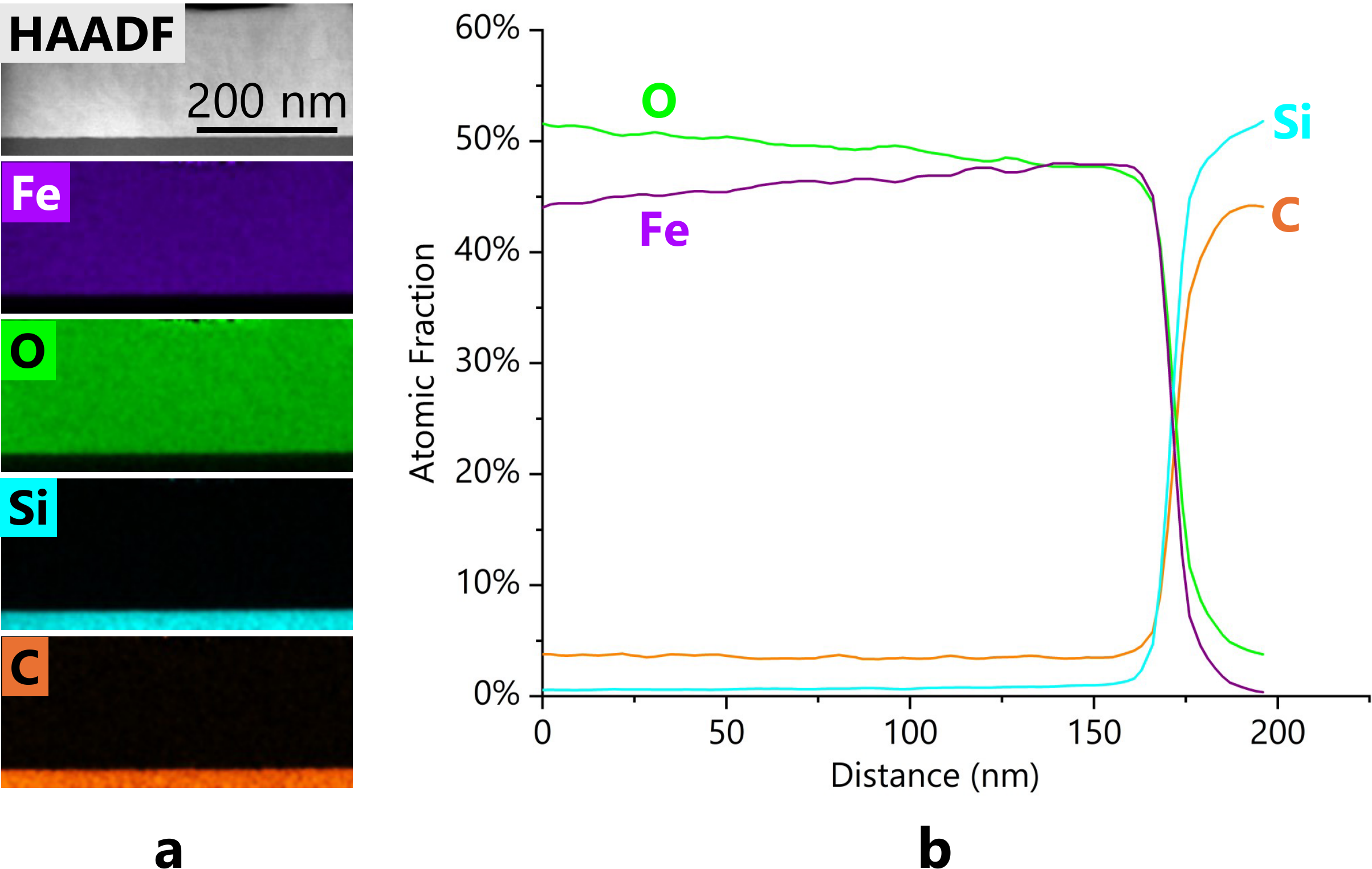} 
\caption{(a) EDS elemental maps of Fe, O, Si, and C taken from Sample D; the region of interest is shown in the included HAADF S/TEM image. (b) Quantitative EDS elemental line profiles, from top surface through the FeO/SiC interface.} 
  \label{fig7} 
\end{figure}

To confirm the composition of the epilayer, EDS elemental analysis was performed, as shown in Figure \ref{fig7}. The sharp FeO/SiC interface can again be easily seen within the qualitative elemental maps in Fig. \ref{fig7}(a). Figure \ref{fig7}(b) presents quantitative elemental concentration profiles for Fe, O, Si, and C. Highly accurate quantification with EDS is only possible with the use of appropriate calibration standards; lighter elements, like O and especially C, are even more difficult due to their low sensitivity. As such, we estimate uncertainties of up to 5$\%$ absolute for the Fe and O profiles. Nonetheless, any trends observed are expected to be at least qualitatively reliable. To that end, it is interesting to find a slight divergence of the Fe and O fractions, ranging from nominally equal (stoichiometric) near the interface to a non-stoichiometric composition of Fe$_{0.85}$O at the epilayer surface. While these exact compositions may not be accurate, the nearly linear compositional non-uniformity is consistent with the broadening observed in the XRD analyses. The source of this non-uniformity is unknown, although it is possible that either the Fe or O$_2$ (or both) fluxes drift slightly during the growths. Similarly, we note that the samples were stored in air for multiple weeks prior to the extraction of the FIB specimen, and the lamella itself was only stored in a plain air desiccator between multiple imaging sessions; thus, additional oxidation of both the film and lamella surfaces cannot be excluded. Regardless, even the highest O concentration indicated by the EDS falls well short of the stoichiometry for magnetite Fe$_3$O$_4$ (i.e., Fe$_{0.75}$O), further supporting the conclusion of effectively phase pure wüstite FeO.

Further investigation of the chemical nature and phase purity of the epilayer was performed using core-loss EELS, focusing on the O ${K}$ and Fe ${L_{2,3}}$ edges. Technically, both the Fe ${L_{2,3}}$ edge and the near-edge fine structure provide information related to the oxidation state of the ion. However, the Fe$^{2+}$ $\leftrightarrow$ Fe$^{3+}$ chemical shift is relatively small at only $\sim$2 eV, and the near-edge structure differences (e.g. white line intensity ratios) are subtle and do not have well-established reference values, ultimately making Fe ${L_{2,3}}$ core-loss analysis difficult \cite{EELs_Colliex_1, EELs_max}. Fortunately, the O ${K}$ edge provides similar information on the Fe oxidation state and bonding environment of the compound, with a very distinct line shape difference between Fe$^{2+}$ and Fe$^{3+}$. In iron oxide compounds where Fe(III)-O bonding is present, like in Fe$_3$O$_4$ (magnetite) and Fe$_2$O$_3$ (hematite), the O ${K}$ edge exhibits a strong pre-peak located at $\sim530 eV$, about $10 eV$ separated from the principal peak; in FeO, where only Fe(II)-O bonding is present, the pre-peak appears only as a minor shoulder feature  \cite{EELs_Colliex_1, EELs_Bischoff_2, mICHEL_3, Chen_6}. 

\begin{figure}[t!]
  \includegraphics[width=0.8\columnwidth]{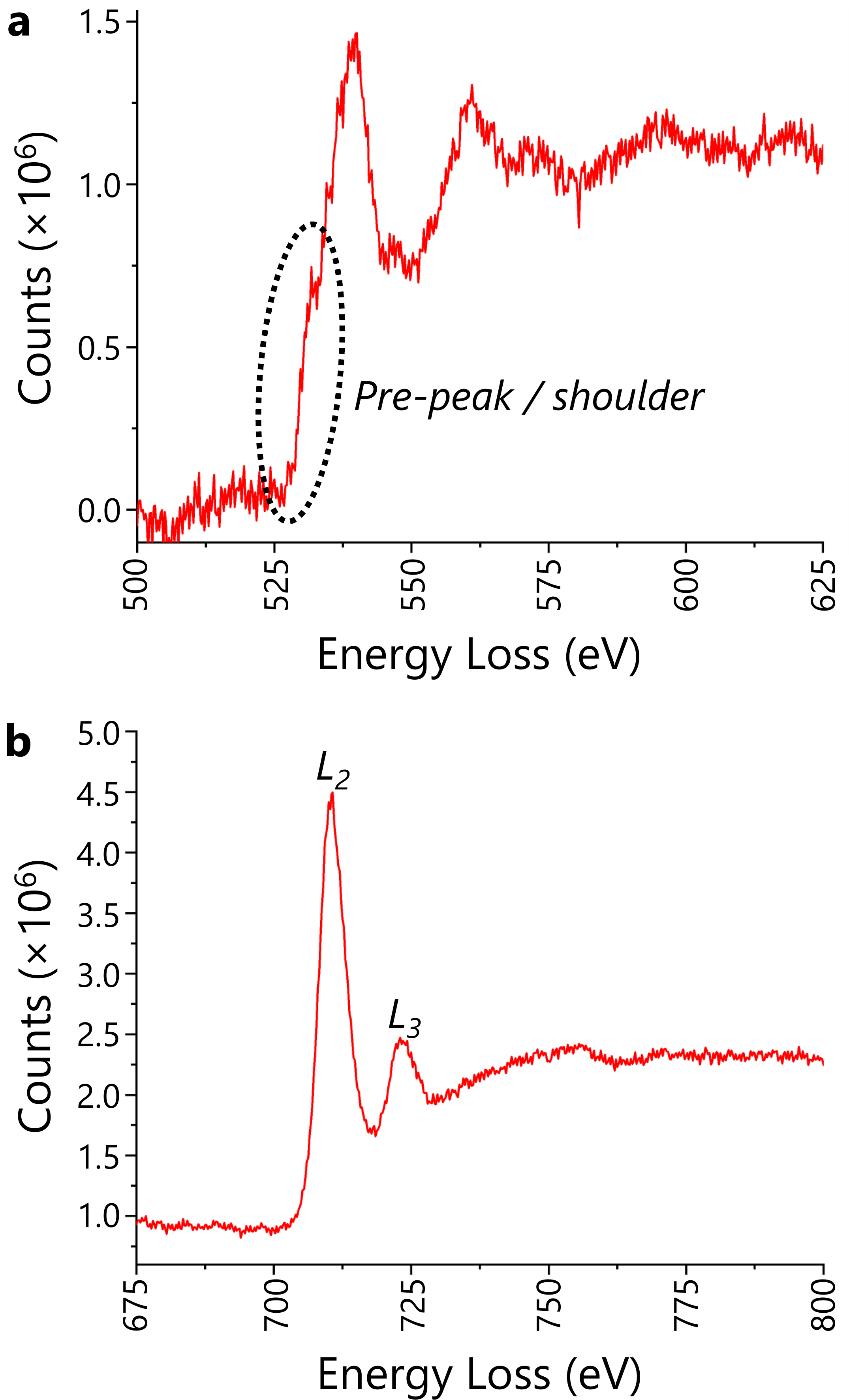} 
\caption{Core-level EELS spectra, collected from and averaged across the FeO (Sample D) epilayer region indicated in Fig. 7(a), of the (a) oxygen $K$ and (b) iron $L_{2,3}$ edges.} 
  \label{fig8} 
\end{figure}

To that end, Figures \ref{fig8}(a) and \ref{fig8}(b) present representative EELS O ${K}$ and Fe ${L_{2,3}}$ core-loss spectra, respectively, acquired over the area indicated in Fig. \ref{fig7}(a) from the Sample D lamella. Both spectra have undergone standard power law background subtraction. Here we focus on analysis of only the O ${K}$ edge spectrum, but the Fe ${L_{2,3}}$ spectrum is provided for completeness. The O ${K}$ edge spectrum in Fig. \ref{fig8}(a) does not exhibit a distinct pre-peak, instead only possessing a weak shoulder at $\sim532 eV$, identical to that reported for Fe(II)-O bonding in pure FeO \cite{EELs_Colliex_1}. Furthermore, upon comparison of the energy-loss near-edge fine structure of the O ${K}$ edge with established reference spectra for various iron oxides \cite{EELs_Colliex_1, EELs_Bischoff_2, mICHEL_3, Garvie_4, wang_5, Chen_6}, the sample is found to match, in both line shape and peak positions, those of wüstite FeO, with no evidence of any appreciable Fe$^{3+}$ content. 

\section{CONCLUSION}

Employing the concept of epitaxial stabilization, the thermodynamically unstable form of iron(II) oxide, FeO (wüstite), was successfully grown on $(0001)$-oriented 4H-SiC via molecular beam epitaxy. Using a combination of pre-growth substrate preparation to establish the desired heterovalent O-bridge interfacial chemistry, growth conditions that yielded the appropriate stoichiometry (and avoided formation of the more stable metallic $\alpha$-Fe or magnetite Fe$_3$O$_4$ phases), and relatively fast quenching to room temperature to help prevent phase separation during cooldown. The FeO heteroepitaxial films, ranging in thickness up to 180 nm, an order of magnitude thicker than any previously reported, were confirmed to be phase-pure wüstite based on a range of structural (XRD, S/TEM) and spectroscopic (EDS, EELS) characterization modalities. XRD RSM analysis using a combination of symmetric and asymmetric diffraction conditions reveals a slight misfit strain-induced rhombohedral distortion of the cubic (halite) crystal structure. These results demonstrate the power of epitaxial stabilization for integrating a thermodynamically unstable, yet functionally interesting (antiferromagnetic FeO) material with a commercially available and technologically important semiconductor platform (4H-SiC).

\begin{acknowledgments}
This materials is based upon work supported by the Air Force Office of Scientific Research (AFOSR) under award number FA9550-23-1-0330.
\end{acknowledgments}

\bibliography{FeO}

\end{document}